\documentclass[aps,pra,twocolumn,groupedaddress,showpacs]{revtex4-1}
\usepackage{amssymb,amsmath,amsfonts,graphicx,latexsym,bm,color}
\usepackage{hyperref}

\newcommand{\beq}{\begin{eqnarray}}
\newcommand{\eeq}{\end{eqnarray}}
\newcommand{\bem}{\begin{pmatrix}}
\newcommand{\eem}{\end{pmatrix}}
\newcommand{\nn}{\nonumber}
\newcommand{\hb}{\hbar}

\newcommand{\f}{\frac}

\newcommand{\dr}[1]{|#1\rangle}
\newcommand{\dl}[1]{\langle#1}
\newcommand{\tb}[1]{\textbf{#1}}
\newcommand{\tr}[1]{\textrm{#1}}

\def\nn{\nonumber}

\def\o{\omega}

\begin{document}

\title{Internal Josephson Oscillations for Distinct Momenta Bose-Einstein Condensates}
\author{Lih-King Lim$^{1,2}$, T. Troppenz$^1$, and C. Morais Smith$^1$}
\affiliation{\centerline{$^1$Institute for Theoretical Physics, Utrecht University, Leuvenlaan 4, 3584 CE Utrecht, The Netherlands}\\
\centerline{$^2$Laboratoire de Physique des Solides, CNRS UMR 8502, Univ. Paris-Sud, F-91405 Orsay cedex, France}}
\date{\today}

\begin{abstract}
The internal Josephson oscillations between an atomic Bose-Einstein condensate (BEC) and a molecular one are studied for atoms in a square optical lattice subjected to a staggered gauge field. The system is described by a Bose-Hubbard model with complex and anisotropic hopping parameters that are different for each species, i.e., atoms and molecules. When the flux per plaquette for each species is small, the system oscillates between two conventional zero-momentum condensates. However, there is a regime of parameters in which Josephson oscillations between a vortex-carrying atomic condensate (finite momentum BEC) and a conventional zero-momentum molecular condensate may be realized. The experimental observation of the oscillations between these qualitatively distinct BEC's is possible with state-of-the-art Ramsey interference techniques.
\end{abstract}

\maketitle

\date{\today}

\section{Introduction}
Quantum coherence is a subject of fundamental importance and practical interest, especially
concerning the construction of quantum logic devices. The Josephson effect, historically proposed to occur in superconductors, has become an important tool for quantum coherence  measurements. In cold atoms, a quantum superposition between two chemically different species (atoms and molecules), which is yet another ramification of the same effect, has been observed by means of Ramsey-like interference experiments for Bose-Einstein condensates (BECs) in a trap \cite{Donley:02,Kokkelmanns,Rembert}. Later, experiments and theoretical studies included an optical lattice \cite{Syassen:07,Dennis,Bhaseen:09}. However, the studies were restricted to relatively simple quantum systems.

The use of Berry's phase to realize artificial gauge fields for cold atoms has proven to be very fruitful to emulate more {\it complex} quantum systems \cite{Dalibard:09}. The possibility of creating vortex lattices with this technique \cite{Lin:09} provides an exciting prospect of reaching the fractional quantum Hall regime. Besides, the generation of a
staggered magnetic flux in a driven two-dimensional (2D) optical lattice holds promises of unprecedented simplicity \cite{Andi,Lih1,Lih2,Cooper,Cooper:11,Kapit:11,Aidelsburger:11}. By loading a staggered-flux optical lattice with cold bosonic atoms, distinct superfluids can form,  depending on the value of the flux $\phi$ per plaquette.
For $\phi < \pi$, the bosons condense at zero momentum, whereas for $\phi > \pi$ a finite momentum BEC is realized, which carries a vortex-antivortex lattice \cite{Lih1,Moller}.
In addition, when manipulating the interactions in the system by means of a Feshbach resonance, a bound state of two bosonic atoms (a Feshbach molecule) can occur, thus raising even further the parameter space for the realization of different BECs: indeed, each type of particles, atoms $(\sigma=1)$ and  molecules $(\sigma=2)$ can, in principle,  condense either at zero or at finite momentum.

Here, we first study the two-body problem of the 2D staggered-flux lattice for cold bosonic atoms.
The staggered flux splits the lattice into ${\cal A}$ and ${\cal B}$ sublattices, thus introducing a  pseudospin degree of freedom into the problem. Due to the breaking of time-reversal symmetry and the presence of the pseudospin, bound states always appear, irrespective of the repulsive interaction strength.
This surprising finding provides a unique opportunity for realizing cold atoms experiments in which pseudospin degrees of freedom play an important role in scattering processes. They may also shed light on the corresponding fermionic many-body problem in the context of high-$T_c$ cuprates. Indeed, a staggered-$\pi$-flux phase was first proposed by Marston and Affleck \cite{Marston} to describe the pseudogap regime of high-$T_c$ cuprates, and it has been advocated by many to be the hidden-order behind high-$T_c$ superconductivity \cite{Laughlin}.

Second, by taking into account a molecular formation, we use the
Bogoliubov theory to study the collective behavior of the generalized
Bose-Hubbard model at zero temperature. The interplay between species and pseudospin
degrees of freedom results in an effective four-band description that
supports various out-of-phase collective modes, also known as Leggett
modes \cite{Leggett:66}. In particular, we find a regime of parameters
in which coherent oscillations between two qualitatively distinct BECs
can be realized: When the flux per plaquette for each species
$\phi_{\sigma=1,2} < \pi$ and the hopping amplitudes $J_{\sigma=1,2} > 0$, the
system oscillates between two conventional zero-momentum condensates.
However, for $\pi < \phi_{\sigma=1} < 3 \pi $, $3\pi < \phi_{\sigma=2} < 4 \pi $, and
$J_{\sigma=2} < 0$, a coherent oscillation between a conventional
zero-momentum molecular BEC and a vortex-antivortex carrying atomic
BEC can occur. The latter describes an internal Josephson oscillation
between two macroscopic groundstates carrying different quantum
numbers \cite{Leggett:01}.

The outline of this paper is the following: in Sec. II we introduce the model and calculate
the two-atom scattering, the many-body problem, and the collective modes in Subsecs. IIA, IIB, and IIC,
respectively. In Sec. III we provide estimates for the realization of the collective modes. In Sec. IV we discuss the possibilities of observing our results experimentally, and we present our conclusions.

\section{The Model}
\begin{figure}
\includegraphics[scale=0.5, angle=0, origin=c]{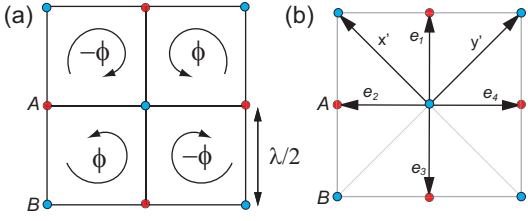}
\caption{\label{Fig.1}(Color online) Optical square lattice in the presence of an artificial staggered flux. (a) Schematic representation of a staggered flux pattern with strength $\phi$ in each plaquette.
(b) Two interpenetrating square sublattices  $\mathcal{A}$ and $\mathcal{B}$ connected by the vectors $\mathbf{e}_l,l\!\in\!\{1,2,3,4\}$. The unit cell is a square lattice (red circles) with a lattice constant $a$.}
\end{figure}

We consider a 2D Bose-Hubbard model for single-species cold bosonic atoms in the presence of a staggered flux,
\begin{eqnarray} \label{Ham1}
H&=&-\sum_{\mathbf{r}\in\mathcal{A},l} \bigl( J e^{i\phi(-1)^l/4} a^{\dag}_{\mathbf{r}} b_{\mathbf{r}+\mathbf{e}_l}+\tr{h.c.}\bigr)\nn\\&&+\f{U}{2}\sum_{\mathbf{r} \in \mathcal{A}\oplus \mathcal{B}} n_\mathbf{r}(n_\mathbf{r}-1).
\end{eqnarray}
Here, $J$ is the hopping amplitude between nearest-neighbor sites, $\phi$ is the flux per plaquette, which alternates in sign between neighboring plaquettes, $l\!\in\!\{1,2,3,4\}$,
and $U$ is the on-site repulsive Hubbard interaction. The operators $a^\dagger_{\mathbf{r}}  (a_{\mathbf{r}})$ and  $b^\dagger_{\mathbf{r}} (b_{\mathbf{r}})$ are the bosonic creation (annihilation) operators at site $\mathbf{r}$ in the square sublattices $\mathcal{A}$ and $\mathcal{B}$, respectively, and $n_\mathbf{r}$ is the number operator. The sublattice constant is given by $a=\lambda/\sqrt{2}$, where $\lambda$ is the laser wavelength and the connecting vectors $\mathbf{e}_l$ are defined by $\mathbf{e}_1=-\mathbf{e}_3=(\lambda/2)\,\mathbf{\hat{e}}_x$ and $\mathbf{e}_2=-\mathbf{e}_4=(\lambda/2)\,\mathbf{\hat{e}}_y$; see Fig.~\ref{Fig.1}. With the techniques described in Ref.~\cite{Andi}, a \textit{strongly} driven time-dependent optical lattice can give rise to an effective hopping amplitude $J$ for the Hamiltonian~(\ref{Ham1}), which takes positive as well as negative values \cite{Andi2}. A physical estimation of the parameters based on Ref.~\cite{Andi} will be discussed in Sec. III.

\subsection{Two-atom scattering}
\begin{figure}
\includegraphics[scale=0.42, angle=0, origin=c]{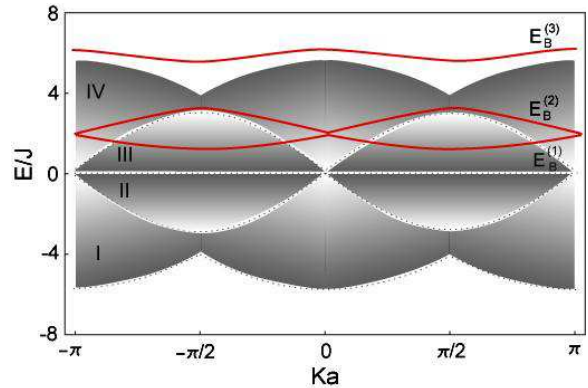}
\caption{\label{Fig.2}(Color online) Energy spectrum $E$ (in units of the hopping parameter $J$) of the two-body Hamiltonian (\ref{Ham1}) as a function of the center-of-mass quasimomentum $\tb{K}$ for $U/J=1.818$ and $\phi=\pi$. The four shaded bands I-IV depict the scattering continua for the four pseudospin components, and the solid curves are the energy dispersion of the repulsively bound pairs that lie above their respective scattering continuum.}
\end{figure}

Let us start by considering the scattering of two bosonic atoms at lattice sites
${\bf r}_1$ and ${\bf r}_2$ of the square optical lattice. The two-body Schr\"odinger equation related to the Hamiltonian (\ref{Ham1}) is given by
\beq \label{Ham2}
\biggl[ H_{{\bf r}_1}\otimes {\mathbb I}_{2\times2}+{\mathbb I}_{2\times2}\otimes H_{{\bf r}_2}+ U\, \delta_{\tb{r}_1,\tb{r}_2} \biggr]{\vec \Psi}=E{\vec \Psi}
\eeq
where
\beq
H_{{\bf r}_i}\equiv\bem
0 & \Delta_{{\bf r}_i}\\
\Delta^*_{{\bf r}_i}&0
\eem
\eeq
is the single-particle kinetic term for particle $i=1,2,$ with the discrete displacement operator $\Delta_{{\bf r}}$ in the presence of staggered flux $\phi$ defined as
\beq
\Delta_{\tb{r}}\Psi(\tb{r})&=&-J \biggl\{e^{i\phi/4}[ \Psi(\tb{r}+\mathbf{e}_1)+\Psi(\tb{r}+\mathbf{e}_3)]\nn\\
&&+e^{-i\phi/4}[\Psi(\tb{r}+\mathbf{e}_2)+\Psi(\tb{r}+\mathbf{e}_4)]\biggr\},\nn\\
\Delta_{\tb{r}}^*\Psi(\tb{r})&=&-J \biggl\{e^{-i\phi/4}[ \Psi(\tb{r}+\mathbf{e}_1)+\Psi(\tb{r}+\mathbf{e}_3)]\nn\\
&&+e^{i\phi/4}[\Psi(\tb{r}+\mathbf{e}_2)+\Psi(\tb{r}+\mathbf{e}_4)]\biggr\},
\eeq
and ${\mathbb I}_{2\times2}$ is the $2\times 2$ identity matrix. The two-particle wavefunction $\vec \Psi$ is constructed by taking the tensor product of two spinorial wavefunctions $(\Psi_{\mathcal{A}}(\tb{r}_1),\Psi_{\mathcal{B}}(\tb{r}_1))\otimes (\Psi_{\mathcal{A}}(\tb{r}_2),\Psi_{\mathcal{B}}(\tb{r}_2))$, or, in terms of its component,
$${\vec \Psi}^T \equiv [ \Psi_{\mathcal{AA}}(\tb{r}_1,\tb{r}_2), \Psi_{ \mathcal{AB}}(\tb{r}_1,\tb{r}_2), \Psi_{\mathcal{BA}}(\tb{r}_1,\tb{r}_2), \Psi_{ \mathcal{BB}}(\tb{r}_1,\tb{r}_2)].$$ Similarly Eq.~(\ref{Ham2}) can be written explicitly as
\beq
\bem
U\delta_{\tb{r}_1,\tb{r}_2}&\Delta_{\tb{r}_2}&\Delta_{\tb{r}_1}&0\\
\Delta^*_{\tb{r}_2}&U\delta_{\tb{r}_1,\tb{r}_2}&0&\Delta_{\tb{r}_1}\\
\Delta^*_{\tb{r}_1}&0&U\delta_{\tb{r}_1,\tb{r}_2}&\Delta_{\tb{r}_2}\\
0&\Delta^*_{\tb{r}_1}&\Delta^*_{\tb{r}_2}&U\delta_{\tb{r}_1,\tb{r}_2}
\eem
{\vec \Psi}=E{\vec \Psi}.
\label{Ham}
\eeq
By using a unitary matrix \cite{Sabio:10},
\beq
\hat{S}\equiv \bem
1&0&0&0\\
0&\f{1}{\sqrt{2}}&\f{1}{\sqrt{2}}&0\\
0&\f{1}{\sqrt{2}}&-\f{1}{\sqrt{2}}&0\\
0&0&0&1\eem,
\eeq
Eq.~(\ref{Ham}) can be expressed in terms of the center-of-mass $\tb{R} = (\tb{r}_1+\tb{r}_2) /2$ and relative $\tb{r} = \tb{r}_1 - \tb{r}_2$
coordinates,
\begin{widetext}
\beq
\hat{S} H \hat{S}^\dag=\bem
U\delta_{\tb{r}_1\tb{r}_2}&\f{1}{\sqrt{2}} ( \Delta_{\tb{r}_1}+\Delta_{\tb{r}_2})&\f{1}{\sqrt{2}}(\Delta_{\tb{r}_2}- \Delta_{\tb{r}_1})&0\\
\f{1}{\sqrt{2}}(\Delta^*_{\tb{r}_1}+\Delta^*_{\tb{r}_2})&U\delta_{\tb{r}_1\tb{r}_2}&0&\f{1}{\sqrt{2}}(\Delta_{\tb{r}_1}+\Delta_{\tb{r}_2})\\
\f{1}{\sqrt{2}}(\Delta^*_{\tb{r}_2}-\Delta^*_{\tb{r}_1})&0&U\delta_{\tb{r}_1\tb{r}_2}&\f{1}{\sqrt{2}}(\Delta_{\tb{r}_1}-\Delta_{\tb{r}_2})\\
0&\f{1}{\sqrt{2}}(\Delta^*_{\tb{r}_1}+\Delta^*_{\tb{r}_2})&\f{1}{\sqrt{2}}(\Delta^*_{\tb{r}_1}-\Delta^*_{\tb{r}_2})&U\delta_{\tb{r}_1\tb{r}_2}
\eem.
\eeq
\end{widetext}
Using plane-wave states $ \hat{S}{\vec \Psi}(\tb{r}_1,\tb{r}_2)=e^{i\tb{K}\cdot \tb{R}} e^{i\tb{k} \cdot \tb{r}}{\vec \Psi_{\tb{K},\tb{k}}}$ with center-of-mass and relative quasimomenta $\tb{K}$ and $\tb{k}$, respectively,  we
rewrite the non-interacting Schr\"odinger equation, which may then be diagonalized by using another unitary matrix $\hat{S}'$,
to yield
\beq\label{full}
\bigl(E -\lambda_{\tb{K},\tb{k}}^{(i)}\bigr)\Phi^{(i)}_{0,\tb{K},\tb{k}}=0, \tr{\ \ \ for\ \ }i = 1,2,3,4,
\eeq
where $\lambda_{\tb{K},\tb{k}}^{(i)}$ is the $i$-th eigenenergy corresponding to the $i$-th component of the ``pseudospin" eigenvector $\vec{\Phi}_{0,\tb{K},\tb{k}}={\hat S}' \vec{\Psi}_{\tb{K},\tb{k}}$. The subscript zero in the eigenvector $\vec{\Phi}_{0,\tb{K},\tb{k}}$ denotes the non-interacting limit of the problem. Since the interaction matrix $\dl{\tb{k}}|\hat{U}\dr{\tb{k}'}= U \,{\mathbb I}_{4 \times 4}$ is diagonal and momentum-independent, the Lippman-Schwinger equation for the scattering problem,
\beq
\dr{\vec{\Phi}_{\tb{K},\tb{k}}}=\dr{\vec{\Phi}_{0,\tb{K},\tb{k}}}+\hat{G}(E_{\tb{K},\tb{k}}) \hat{U} \dr{\vec{\Phi}_{\tb{K},\tb{k}}}\nn
\eeq
can be resummed to all orders in $\hat{U}$ in the new basis $\vec{\Phi}_{0,\tb{K},\tb{k}}(\tb{r})$ for each pseudospin component to yield
\beq
\Phi_{\tb{K},\tb{k}}^{(i)}(\tb{r})=\Phi^{(i)}_{0,\tb{K},\tb{k}}(\tb{r})+\f{U G^{(i)}(E_{\tb{K},\tb{k}},\tb{r})}{1-U G^{(i)} (E_{\tb{K},\tb{k}},0)}.
\eeq
Here, $G^{(i)}(E,\tb{r})$ is the Fourier transform of the non-interacting Green's function $\tilde{G}^{(i)}(E,\tb{k}')=1/(E-\lambda^{i}_{\tb{K},\tb{k}'}+i0)$. In contrast to the usual atomic scattering problem, the pseudospin-carrying particles give rise to four independent scattering continua, see Fig.~\ref{Fig.2}. Each scattering continuum, in addition, may support a bound state lying above it, which occurs when the scattering amplitude diverges, i.e., at the position of the pole with energy $E_{B}^{(i)}$,
\beq
\int_{1BZ} \f{d^2\tb{k}'}{(2\pi)^2}\f{1}{E_B^{(i)}-\lambda_{\tb{K},\tb{k}'}^{(i)}}=\f{1}{U},
\eeq
see solid curves in Fig.~\ref{Fig.2}.

We observe two striking features in the present model, which hold for all flux values except for the cases $\phi=2\pi n$, $n\in \mathbb{Z}$, when the two sublattices become degenerate. Firstly, the existence of at least one flat band $\lambda_{\tb{K}=0,\tb{k}}^{(i)}=0$, which is independent of the relative quasimomentum $\tb{k}$ at zero center-of-mass quasimomentum $\tb{K}=0$, guarantees the existence of a bound state solution with energy $E_B=U$, even for arbitrarily weak repulsive interaction strength. This is in stark contrast to the usual Hubbard model in a simple 3D cubic lattice, where a critical repulsive interaction strength is required for the existence of a bound state solution \cite{Zoller}. Secondly, at certain energies both scattering and on-resonance processes can take place simultaneously in separate pseudospin channels. Collisions between free atoms and repulsive bound pairs can thus lead to interesting dynamics since the pseudospin is generally not a good quantum number. Moreover, due to the lack of dissipation in an optical lattice, the repulsive bound pairs are expected to be long-lived \cite{Zoller}, and it is conceivable that they can form a BEC.

\subsection{Many-body problem}
To study the dynamics when both atoms and composite particles are present, we generalize the Hamiltonian (\ref{Ham1}) to an effective two-species model:
\beq
H =&-& \sum_{\mathbf{r}\in\mathcal{A},l,\sigma}\bigl(J_\sigma e^{i\phi_{\sigma}(-1)^l/4} a^{\dag}_{\mathbf{r},\sigma} b_{\mathbf{r}+\mathbf{e}_l,\sigma}+\tr{h.c.}\bigr)  \nn \\
&+&  \sum_{\mathbf{r} \in \mathcal{A}\oplus \mathcal{B},\sigma}\biggl[
(\epsilon_\sigma-\mu_\sigma)n_{\mathbf{r},\sigma}+\frac{U_\sigma}{2} n_{\mathbf{r},\sigma}(n_{\mathbf{r},\sigma}-1) \biggr]\nn  \\
&+&g\biggl[  \sum_{\mathbf{r} \in \mathcal{A}} a_{\mathbf{r},2}^\dag a_{\mathbf{r},1} a_{\mathbf{r},1}+\sum_{ \mathbf{r} \in \mathcal{B}} b_{\mathbf{r},2}^\dag b_{\mathbf{r},1} b_{\mathbf{r},1}  + \rm{h.c.} \biggr],
\label{H2}
\eeq
with species index $\sigma=1(2)$ denoting atoms (composite particles), and an inter-species conversion term characterized by the strength $g$. Here, $\epsilon_{\sigma=1(2)}$ and $\mu_{\sigma=1(2)}$ are respectively the on-site energy and the chemical potential for atoms (composite particles), and $\mu_2= 2\mu_1$. The hopping amplitude $J_\sigma$, the flux strength $\phi_\sigma$, and the on-site interaction $U_\sigma$ are species dependent because of the difference in masses $m_2\!=\!2 m_1$. Multi-band effects are neglected because only small inter-atomic interactions are considered.
In addition to repulsive bound pairs, composite particles (which we henceforth call molecules) can also be formed by the use of Feshbach resonances \cite{Chin:10}. In the absence of a staggered flux, this two-species model can in fact be derived from a microscopic Feshbach resonance model in which the relative on-site energy $\epsilon_{\sigma=1}/\epsilon_{\sigma=2}$ and the inter-species coupling strength $g$ can be tuned via an external Feshbach magnetic field \cite{Dennis,Diener:06,Buchler:10}. The molecule-molecule interaction $U_2$ and the atom-molecule interaction terms are neglected in the present work because they are of higher order in the molecule operators, and we are considering the limit of a low molecule filling factor.
Throughout our work, we consider $n_2 \ll\, 0.3$ and these terms have a negligible contribution.

\begin{figure}
\includegraphics[scale=1, angle=0, origin=c]{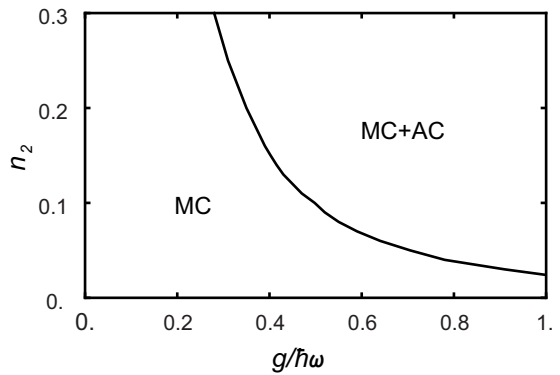}
\caption{\label{Fig.3}(Color online) The phase boundary of the quantum Ising transition plotted as a function of the molecular condensate fraction $n_2$ and the coupling strength $g/\hb \omega$. Each lattice site represents a sufficiently deep harmonic trap with frequency $\hb \omega=10 \tr{kHz}$, in terms of which all energy units are based. The figure is plotted for the following chosen parameters: $J_1/\hb \omega=-0.11, J_2/\hb \omega=-0.08, U_1/\hb \omega=0.2, \epsilon_1/\hb \omega=1.5, \epsilon_2/\hb \omega=2, \phi_1=4\pi/3$, and $\phi_2=16\pi/5$.}
\end{figure}

The corresponding phase diagram exhibits a rich behavior, which includes an array of superfluid-Mott-insulator transitions \cite{Bhaseen:09} and a quantum Ising transition due to an enlarged symmetry group $U(1)\times Z_2$ \cite{Radzihovsky:04,Romans:04}. The Ising transition, which can also be viewed as a topological confinement-deconfinement transition, separates a phase with a molecular condensate (MC) and normal atoms with a residual $Z_2$ symmetry, from a phase in which both molecules and atoms (MC+AC) are condensed. The mean-field phase boundary can be determined by generalizing the work of Ref.~\cite{Dennis} and identifying the development of an instability in the atomic spectrum in the phase with only a molecular condensate; see Fig.~\ref{Fig.3}. We will henceforth measure the energy parameters in units of $\hbar \omega=10 \tr{kHz}$, which is the typical harmonic frequency of each well on a lattice site. At higher values of the interaction strength, the staggered flux is expected to simply renormalize the critical value $(U/4J)_c$ of the superfluid-Mott-insulator transition at a fixed density, as it was found for single-species \cite{Lih1}.

\begin{figure}
\includegraphics[scale=0.3, angle=0, origin=c]{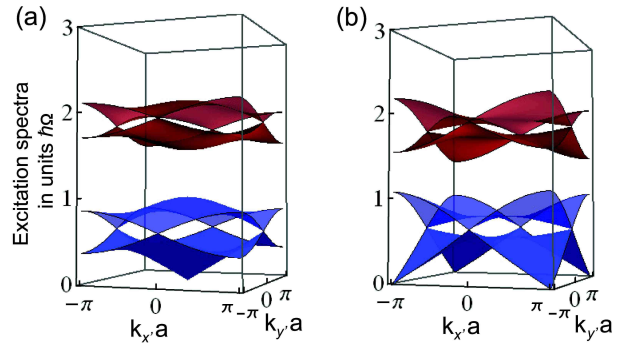}
\caption{\label{Fig.4}(Color online) Collective modes of the atom-molecule system subjected to a staggered flux: (a) Scenario (I) with $J_1/\hb \omega=0.11, J_2/\hb \omega=0.08$, and $\phi_{1,2}=4\pi/5$. (b) Scenario (II) with $J_1/\hb \omega=-0.11, J_2/\hb \omega=-0.08, \phi_1=4\pi/3$, and $\phi_2=16\pi/5$.  The other parameters are chosen as follows: $g/\hb \omega=0.8, U_1/\hb \omega=0.2, \epsilon_1/\hb \omega=1.5, \epsilon_2/\hb \Omega=2,$ $n_1=0.5$, and  $\hb \omega=10 \tr{kHz}$.}
\end{figure}

\subsection{Collective modes}
Deep in the superfluid regime, however, BEC states with distinct momenta can concurrently form because the atomic and molecular fields are subjected to different flux values $\phi_{\sigma}$. In this subsection, we focus on the collective modes in the superfluid regime where the $U(1)\times Z_2$ symmetry is broken. By performing a canonical transformation to the band representation,
\beq
\alpha_{\tb{k},\sigma}^+&=&\f{1}{\sqrt{2}}\biggl(\f{\tilde{\epsilon}_{\tb{k},\sigma}}{|\tilde{\epsilon}_{\tb{k},\sigma}|} a_{\tb{k},\sigma}+b_{\tb{k},\sigma}\biggr),\\
\alpha_{\tb{k},\sigma}^-&=&\f{1}{\sqrt{2}}\biggl( -\f{\tilde{\epsilon}_{\tb{k},\sigma}}{|\tilde{\epsilon}_{\tb{k},\sigma}|} a_{\tb{k},\sigma}+b_{\tb{k},\sigma}\biggr),\nn
\eeq
where
\beq
\tilde{\epsilon}_{\tb{k},\sigma}&=&4J_\sigma \biggl[ \cos\biggl(\f{\phi_\sigma}{4}\biggr) \cos\biggl(\f{k_{x'} a}{2}\biggr) \cos\biggl(\f{k_{y'} a}{2}\biggr)\nn\\
&&-i \sin\biggl(\f{\phi_\sigma}{4}\biggr) \sin\biggl(\f{k_{x'} a}{2}\biggr) \sin\biggl(\f{k_{y'} a}{2}\biggr) \biggr]
\eeq
is the kinetic form factor in the momentum space (the prime denotes sublattice coordinates), we find that the non-interacting part of the Hamiltonian~(\ref{H2}) is described by four distinct bands with energy dispersions
\beq
E_{\tb{k},\sigma}^\pm&=&\pm2J_{\sigma} \biggl\{\cos^2[k^+ a]+\cos^2[k^- a]\nn\\
&&+2 \cos[ \phi_\sigma/2] \cos[k^+ a]\cos[k^- a]\biggr\}^{1/2},
\eeq
where $k^{\pm}=(k_{x'}\pm k_{y'})/2$ (see Fig.\ 2 for the definition of the $x'-y'$ coordinate system). The on-site interaction and atom-molecule coupling terms take the general inter-/intra-band coupling form. We then consider the formation of BECs, where the lowest energy operators for the atoms $(\sigma=1)$ and for the molecules $(\sigma=2)$, both acquire a non-zero expectation value at the respective condensation points $k_0(\sigma=1,2)$, with condensate number $N_{0,\sigma}$. The Bogoliubov approximation amounts to substituting $\alpha^{-}_{k=k_0(\sigma),\sigma}\!\rightarrow\! \sqrt{N_{0,\sigma}}+\alpha^{-}_{k=k_0(\sigma),\sigma}$ and keeping the fluctuations up to the quadratic order.
The $8\times8$ grand-canonical mean-field Hamiltonian then reads
\beq\label{MFHam}
H=\f{1}{2}\sum_{i\omega_n, \tb{k}}\hat{\psi}^\dag_{\tb{k}} \hat{G}_{\tb{k}} \hat{\psi}_{\tb{k}}, \,\,
\hat{G}_{\tb{k}}=\left[ \begin{array}{cc}\hat{M}_{\tb{k},1}&\hat{Q}_{\tb{k}}\\\hat{Q}_{\tb{k}}^\dag&\hat{M}_{\tb{k},2}
\end{array}
\right],
\eeq
where the matrices
\beq
\hat{Q}_{\tb{k}}=g\left[ \begin{array}{cccc}\sqrt{n_1}(s^-_k)^*&0&\sqrt{n_1}(s^+_k)^*&0\\0&\sqrt{n_1}s^-_{-k}&0&\sqrt{n_1}s^+_{-k}\\\sqrt{n_1}(s^+_k)^*&0&\sqrt{n_1}(s^-_k)^*&0\\0&\sqrt{n_1}s^+_{-k}&0&\sqrt{n_1}s^-_{-k},\end{array}
\right], \nn
\eeq
\begin{widetext}
\beq
&\hat{M}_{k,1}=\left[ \begin{array}{cccc}\varepsilon_{k,1}^+&\f{1}{2}U_{1} n_1 u^+_k +g\sqrt{n_2}(v^-_k)^*&0&\f{1}{2}U_1 n_1 u^-_k+g\sqrt{n_2}(v^+_k)^*\\
\f{1}{2}U_{1} n_1 (u^+_k)^*+g\sqrt{n_2}v_{k}^-&(\varepsilon_{k,1}^+)^*&\f{1}{2}U_1 n_1 (u^-_{-k})^*+g\sqrt{n_2}v^+_{-k}&0\\0&\f{1}{2}U_1 n_1 u^-_{-k}+g\sqrt{n_2}(v^+_{-k})^*&\varepsilon_{k,1}^-&\f{1}{2}U_1 n_1 u^+_k+g\sqrt{n_2}(v_k^-)^*\\
\f{1}{2}U_1 n_1 (u^-_k)^*+g\sqrt{n_2}v^+_{k}&0&\f{1}{2}U_1 n_1 (u^+_k)^*+g\sqrt{n_2}v^-_{k}&(\varepsilon_{k,1}^-)^*,\end{array}
\right],& \nn\\
&\hat{M}_{k,2}=\left[ \begin{array}{cccc}\varepsilon_{k,2}^+&0 &0&0\\
0&(\varepsilon_{k,2}^+)^*&0&0\\0&0&\varepsilon_{k,2}^-&0\\
0&0&0&(\varepsilon_{k,2}^-)^*,\end{array}
\right],&\nn
\eeq
with the Nambu spinor defined as
\beq
\hat{\psi}^\dag_{k}=\biggl(\alpha^{+\,\,\dag}_{k,1},\,\alpha^{+}_{-k,1},\,
\alpha^{-\,\,\dag}_{k,1},\,\alpha^{-}_{-k,1},\,\alpha^{+\,\,\dag}_{k+k_0(\sigma=1),2},\,\alpha^{+}_{-k+k_0(\sigma=1),2},\,
\alpha^{-\,\,\dag}_{k+k_0(\sigma=1),2},\,\alpha^{-}_{-k+k_0(\sigma=1),2}\biggr).\nn \eeq
Here, $n_1 (n_2)$ denotes the atomic (molecular) condensate fraction, $\varepsilon_{k,\sigma}^\pm=i\hb\o_n+E_{k,\sigma}^\pm+\epsilon_{\sigma}- \mu_{\sigma}+ 2 U_{\sigma} n_{\sigma}$, where $\omega_n$ is the Matsubara frequency. The transformation coefficients are given as: $u^{\pm}_k=1\pm\eta_{k,1}^* \,\eta_{-k,1}^*\,\eta_{k_0(\sigma=1),1}\,\eta_{k_0(\sigma=1),1}$, $v_k^{\pm}=1\pm\eta_{0,2}^*\,\eta_{k,1}\,\eta_{-k,1}$, $s_k^{\pm}=1\pm\eta_{k+k_0(\sigma=1),2}^*\,\eta_{k_0(\sigma=1),1}\,\eta_{k,1}$, where $\eta_{k,\sigma}=\tr{Arg}\bigl\{e^{i\phi_\sigma/4}\cos[(k_{x'}+k_{y'})a/2]+ e^{-i\phi_\sigma/4}\cos[(k_{x'}-k_{y'})a/2]\bigr\}$ and $k_0(\sigma)$ denotes the quasimomentum at which species $\sigma$ condenses.
\end{widetext}

\begin{figure}
\includegraphics[scale=0.4, angle=0, origin=c]{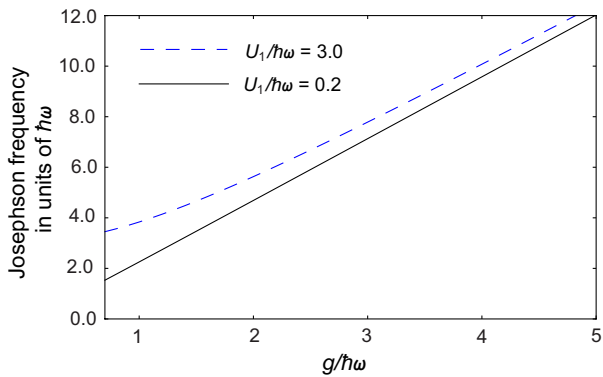}
\caption{\label{Fig.5}(color online) Josephson frequency of two BECs with distinct momenta as a function of coupling strength for different on-site interaction strength, with $\hb \omega=10\tr{kHz}$. The system exhibits dynamical instabilities for $g\lesssim0.7 \hb \omega$.}
\end{figure}
We require that the terms that are linear in the fluctuations vanish, yielding the relations,
\beq
&&\tr{sign}\bigl(J_2\bigr)\exp\bigl(\eta^*_{k_0(\sigma=2),2}+2\eta_{k_0(\sigma=1),1}\bigr)=1,\nn\\
&&\tr{sign}\bigl(J_2\bigr)\exp\bigl(\eta^*_{k_0(\sigma=2),2}+\eta_{k_0(\sigma=1),1}\nn\\
&&\tr{\ \ \ \ \ \ \ \ \ \ \ \ \ \ \ \ \ \ \ \ }+\eta_{k_0(\sigma=2)-k_0(\sigma=1),1}\bigr)=1,
\eeq
as well as
\beq
\label{eqn:linea_term}
&&\biggl[1+\tr{sign}\bigl(J_2\bigr)\,\exp \bigl( \eta^*_{k_0(\sigma=2),2}+2\eta_{k_0(\sigma=1),1} \bigr) \biggr]\nn \\
&&\times g n_1 \sqrt{n_2} \,\delta_{k_0(\sigma=2),2k_0 (\sigma=1)} \nn\\
&&+ n_{\sigma}(1+\delta_{\sigma,2})\,(E_{k,\sigma}^- -\mu_{\sigma}+\epsilon_{\sigma}+U_{\sigma}n_{\sigma})=0 ,
\eeq
where $\delta_{a,b}$ is a Kronecker delta. We find that the two-species system is only allowed to have two condensation
scenarios: (I) for $J_{2}\!>\!0$, $0\!\leqslant\!\phi_{1,2}\!<\!\pi$ and (II)
for $J_{2}\!<\!0$, $\pi\!<\!\phi_{1}\!<\!3\pi$, $3\pi\!<\!\phi_{2}\!<\!4\pi$.
In scenario (I), both atoms and molecules condense at the same momentum $\tb{k}_0=0$,
whereas in scenario (II) the molecules condense at $\tb{k}_0=0$, while the atoms
condense at a different momentum $\tb{k}_0=(\pi/a,\pi/a)$. Outside these regimes, the
system does not admit a self-consistent stable mean-field condensate state.
Since $\mu_2=2\mu_1$, we can equate the two expressions resulting from
solving Eq.~(\ref{eqn:linea_term}) for $\mu_{\sigma}$ and obtain
a relation between $n_1$ and $n_2$, demonstrating that the two condensate
fractions are not independent.

The excitation spectrum for both scenarios, obtained by solving the mean-field Hamiltonian~(\ref{MFHam}), consists of four collective modes because of the $\cal{A\!-\!B}$ sublattices and the atom-molecule species degrees of freedom; see Fig.~(\ref{Fig.4}). The lowest branch is the Goldstone mode associated with both, $\cal{A\!-\!B}$ sublattices and atom-molecule in-phase density fluctuations. The next lowest branch corresponds to the $\cal{A\!-\!B}$ out-of-phase but atom-molecule in-phase oscillation mode. The third and fourth branches, separated by a gap from the two lowest branches, describe collective modes where atoms and molecules oscillate out-of-phase. In fact, the long-wavelength physics of the third branch is equivalent to a coherent oscillation of density number of the two species in real time. It thus describes an internal Josephson effect of an atomic and a molecular condensates that can have distinct quantum numbers, with a frequency that is given by the energy gap. The dependence of the gap on the coupling strength $g$ for scenario (II) is shown in Fig.~(\ref{Fig.5}) for weak and intermediate on-site interactions. This result provides a promising experimental opportunity to study the coherent dynamics of two \textit{distinct} macroscopic ground states in a cold atomic system.

\section{Experimental Observation}

The many-body phenomena discussed in Sec. II.C may be observed experimentally by using the set-up proposed in Refs. \cite{Andi,Andi2}, where ultracold bosons are trapped in a time-dependent square optical  lattice with staggered currents. Here, we provide an estimate of the parameters specific for this configuration. Using the Floquet theory, one can show that the time-dependent problem can be described by an  effective Hamiltonian  of the form \cite{Andi2}
\beq\label{Ham3}
H_{\tr{eff}}&=&-\bar{J} \mathcal{J}_0\biggl( \f{2 \xi_N}{\hb \Omega}\biggr) \hat{T}_+ -i\f{1}{2} \xi_M \mathcal{J}_1\biggl( \f{2 \xi_N}{\hb \Omega} \biggr)[\hat{M},\hat{N}]\nn\\
&&+H_{\tr{int}}\nn\\
&\equiv&-J\, \hat{T}_+-i\f{1}{2}K\,[\hat{M},\hat{N}]+H_{\tr{int}}
\eeq
where $\hat{T}_+\equiv  \sum_{\mathbf{r}\in\mathcal{A},l} \bigl\{a^{\dag}_{\mathbf{r}} b_{\mathbf{r}+\mathbf{e}_l}+\tr{h.c.}\bigr\}$ and $[\hat{M},\hat{N}]\equiv 2 \sum_{\mathbf{r}\in\mathcal{A},l}  \bigl\{(-1)^l a^{\dag}_{\mathbf{r}} b_{\mathbf{r}+\mathbf{e}_l}-\tr{h.c.}\bigr\} $ are the single-particle hopping operators, with $l = 1,2,3,$ and 4. $\mathcal{J}_0(z)$ and $\mathcal{J}_1(z)$ are Bessel functions of the first kind and $H_{\tr{int}}$ is the onsite interaction term. The parameters $\bar{J}=\int dx dy\, w(x,y) w(x+1/2,y)$, $\xi_N=2\kappa \bar{V}_0 \int dx dy \,|w(x,y)|^2 \cos(kx)\cos(ky) $, and $\xi_M=\kappa\, \bar{V}_0\int dx dy\, w^*(x+\lambda/4,y)[\sin^2(kx)-\cos^2(ky)] w(x-\lambda/4,y)$ are given in terms of the Wannier function $w(x,y)$ of the optical lattice with amplitude $\kappa$ and lattice depth $\bar{V}_0$. The time-periodic driving has a frequency $\Omega$, and the \textit{strongly} driven regime is fulfilled by the conditions $\bar{J}, \xi_M,U\ll \hb \Omega$. For simplicity, we may take the Wannier function to be the ground state wavefunction of a harmonic oscillator $w(x,y)=(1/l_{HO}\sqrt{\pi})\exp[-(x^2+y^2)/2 l_{HO}^2]$ for a lattice site with harmonic length $ l_{HO}=(E_r/\bar{V}_0)^{1/4}(\lambda/2\pi )$, where $E_r=h^2 k^2/2m$ is the recoil energy, $m$ is the mass of the atom, and $k = 2 \pi /\lambda$. For an optical lattice potential with amplitude fixed at $\kappa=1$ and $\bar{V_0}/E_r\approx 2$, we get $\xi_N\simeq 1.404 \,\bar{V}_0$. For the other parameters, we use the result from a full band calculation made in Ref. \cite{Andi2} that gives $\bar{J}\simeq 2.5 \,\xi_M$. Because the Bessel functions range from negative to positive values, we see that the tunneling amplitudes $J$ and $K$ can attain negative values as a function of the driving frequency, see Fig. 6(a).

In the superfluid regime of the Hamiltonian~(\ref{Ham3}), there exist four distinct order parameters in four ranges of driving frequency, which we denote  here as superfluid A, B, C and D; see Fig. 6(b): superfluids A and C are the zero-momentum condensate while superfluids B and D are the finite-momentum condensates. As explained in Ref.~\cite{Andi2}, even though two of the four superfluids carry the same momentum, they have different symmetries associated with them. On the other hand, there is a square region enclosing the origin of the phase diagram that characterizes a Mott insulator phase, not shown in Fig. 6(b). For the lattice amplitude we consider here, the width of the Mott region at unit filling is $\simeq 0.04 U$, thus not sizable, and the system does not cross into the Mott insulator state as the driving frequency is varied.

\begin{figure}
\includegraphics[scale=0.7, angle=0, origin=c]{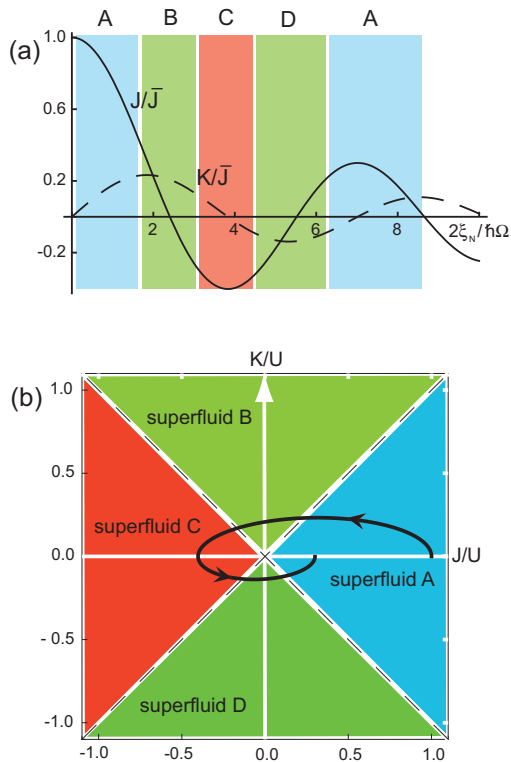}
\caption{(color online) (a) The value of $J/\bar{J}$ (the real component) and $K/\bar{J}$ (the imaginary component) plotted as a function of $2\xi_N/\hb \Omega$, for amplitude $\kappa=1$ and $\bar{V}_0/E_R=2$. The four zones A, B, C and D characterize four superfluids with different symmetries in the weak interaction regime. (b) Phase diagram of the Hamiltonian~(\ref{Ham3}). The solid curve shows the superfluid phases that the system realizes as the parameter $2\xi_N/\hb \Omega$ is tuned from $0$ to $8.5$. We take $U/\bar{J}\approx 1$ for our case.}
\end{figure}

To relate to the Hamiltonian studied in Eq. (1), we may rewrite the kinetic term of Eq. (\ref{Ham3}) in the polar form
\beq
H_{\tr{eff}}&=&-\sum_{\mathbf{r}\in\mathcal{A},l} \bigl( J_{\tr{eff}} e^{i\phi_{\tr{eff}}(-1)^l/4} a^{\dag}_{\mathbf{r}} b_{\mathbf{r}+\mathbf{e}_l}+\tr{h.c.}\bigr)
\eeq
with the effective parameters given as
\beq
J_{\tr{eff}}&=&\sqrt{\bar{J}^2 \mathcal{J}_0^2\biggl( \f{2 \xi_N}{\hb \Omega}\biggr)+\xi_M^2 \mathcal{J}_1^2\biggl( \f{2 \xi_N}{\hb \Omega} \biggr)},\nn\\
\phi_{\tr{eff}}&=& 4 \tan^{-1}\biggl[\f{\xi_M \mathcal{J}_1\bigl( \f{2 \xi_N}{\hb \Omega}\bigr)}{\bar{J} \mathcal{J}_0\bigl( \f{2 \xi_N}{\hb \Omega}\bigr)}\biggr].
\eeq
In this representation, the negativity of the tunneling amplitude arises from the phase value $e^{ i \pi}$.

An additional ingredient of the model we study in this work is that it consists of two species, namely atoms and molecules, trapped in the same time-dependent optical lattice. We take the simplest consideration where their masses are related by $m_2=2\, m_1$. Accordingly, the parameters of the model, through their dependence on the Wannier functions, scale as $l_{HO, \sigma=1}=1.189\, l_{HO, \sigma=2}$, $\bar{J}_{\sigma=1}=3.838\, \bar{J}_{\sigma=2}$,  $\xi_{N,\sigma=1}=0.901 \,\xi_{N,\sigma=2}$, and $\bar{J}_{\sigma=2}\simeq 2.5\, \xi_{M,\sigma=2}$. The additional subscript $\sigma$, as previously used, denotes the species index. Thus, for the molecules subjected to the same time-dependent optical lattice, the positions of the zeros of the tunneling amplitudes $J_{\sigma=2}$ and $K_{\sigma=2}$ (the real and imaginary components, respectively) are both scaled by a factor 0.901 as compared to the atomic case in Fig. 6(a). Consequently, the driving frequency range for the various \textit{molecular} superfluid phases is shifted accordingly.

In Sec. II.C we propose two possible scenarios to study coherent oscillations of atomic and molecular condensates. In the phase diagram of Fig. 6(b), the two scenarios are: (I) both atoms and molecules condense at zero momentum and form the  superfluid A; (II) the atoms condense at finite momentum (superfluid B), whereas the molecules condense at zero momentum (superfluid C). Therefore, for the setup under consideration in this section, we estimate that for a driving frequency range $0 <2\xi_{N,\sigma=1}/\hbar\Omega< 1 $, scenario (I) can be realized, whereas for the driving frequency region  $2\xi_{N,\sigma=1}/\hbar\Omega\approx 3$, scenario (II) can be realized.

\section{Discussions and Conclusions}
Bound states with pseudospin components can be observed experimentally, since the binding energy $E_B^{i}$ can be inferred by spectroscopic measurements \cite{Zoller,Regal:03,Moritz:05}. The lifetime of the bound state can also be probed through the techniques used in Ref.~\cite{Zoller}. More importantly, a coherent oscillation between distinct macroscopic ground states can be experimentally observed by performing double-pulse Ramsey experiments
\cite{Donley:02,Syassen:07} for bosonic atoms in the optical lattice setup described in Ref. \cite{Andi}.
The regime of negative hopping parameters $J < 0$ for the staggered flux optical lattice may be realized experimentally, as shown in Ref. \cite{Andilast}.
Recent experiments in optical lattices without flux have probed the negative hopping regime by shaking the lattice with a periodic perturbation \cite{Arimondo}, similar to the one occurring in Ref. \cite{Andi}.

In conclusion, we investigated the dynamics of a generalized two-species Hubbard model subjected to a staggered flux. At the two-atom level, we studied the scattering problem with an onsite interaction. The local nature of the interaction allowed us to solve the two-body problem exactly, and we obtained the scattering spectra, as well as bound states induced by the repulsive interaction. The fact that the atoms carry a pseudospin gives rise to new scattering properties that are not present in conventional cold atomic systems. In particular, we find that a repulsive bound state always exists at zero center-of-mass quasimomentum due to the existence of a flat band. We then use the existence of bound states in the two-body problem as a motivation to study the many-body problem in a generalized Hubbard model subjected to a staggered flux. The resulting system can execute various collective density oscillations, especially an out-of-phase mode, which is also known as Josephson oscillation. Furthermore, we propose a way to realize a quantum superposition of different species with different momenta, zero for the molecular condensate and finite for the atomic one. The results can be probed by combining state-of-the-art techniques. This work opens up new perspectives in the realization of more complex BEC's and in the precise control of their dynamics.

\begin{acknowledgments}
We acknowledge financial support from the Netherlands Organization for Scientific Research (NWO). We are grateful to A. Hemmerich and I. Spielman for a critical reading of this manuscript.
\end{acknowledgments}

\end{document}